\documentclass[aps,floats,twocolumn,showpacs]{revtex4}
\usepackage[cp1251]{inputenc}
\usepackage[english]{babel}
\usepackage{amssymb}
\usepackage{amsmath}
\usepackage{graphicx}
 \def\ds{\displaystyle}
 \def\bc{\begin{center}}          \def\ec{\end{center}}
 \allowdisplaybreaks

\begin{document}
 \title{Modulational instability of a Langmuir wave in plasmas with energetic tails of superthermal electrons}
 \author{I.V.Timofeev}
 \affiliation{Budker Institute of Nuclear Physics SB RAS, 630090, Novosibirsk, Russia \\
 Novosibirsk State University, 630090, Novosibirsk, Russia}
 \begin{abstract}

The impact of superthermal electrons on dispersion properties of
isotropic plasmas and on the modulational instability of a
monochromatic Langmuir wave is studied for the case when the
power-law tail of the electron distribution function extends to
relativistic velocities and contains most of the plasma kinetic
energy. Such an energetic tail of electrons is shown to increase the
thermal correction to the Langmuir wave frequency, which is
equivalent to the increase of the effective electron temperature in
the fluid approach, and has almost no impact on the dispersion of
ion-acoustic waves, in which the role of temperature is played by
the thermal spread of low-energy core electrons. It is also found
that the spectrum of modulational instability in the non-maxwellian
plasma narrows significantly, as compared to the equilibrium case,
without change of the maximum growth rate and the corresponding
wavenumber.
 \end{abstract}
 \maketitle

\section{Introduction}

The typical feature of turbulence evolution in a beam-plasma system
is the formation of superthermal tails on the distribution function
of plasma electrons. These tails are often observed in space and
laboratory plasmas. In particular, the electron distribution in the
solar wind (see the review paper \cite{pier} and references therein)
was found to be well approximated by the family of kappa-functions
\cite{vas,sum,chat,mace,hel}, decaying in velocity as a power law.
Moreover, it has been shown recently that these functions describe
asymptotically stationary solutions of weak turbulence equations, in
which self-consistent evolution of both wave and particles spectra
is taken into account \cite{yoon1,yoon2,yoon3}. In laboratory
experiments, beam-plasma interaction appears to be much more intense
than in the solar wind, that is why it results in the formation of
strongly non-maxwellian electron distribution, in which most of the
plasma kinetic energy is concentrated in a rather small population
of fast electrons. Such a slowly decaying momentum distribution of
plasma electrons ($f\propto p^{-5}$) was really observed in the
experiments on turbulent plasma heating by the powerful electron
beam in the multi-mirror trap GOL-3 \cite{burd}.

It is obvious that such an energetic tail of superthermal electrons
should modify not only the linear dispersion of plasma modes, but
also probabilities of various nonlinear processes responsible for
formation of the turbulent spectrum. In the experiments of interest
\cite{arz}, the high-current relativistic electron beam interacts
with the plasma in the regime of strong plasma turbulence. In this
case, the main nonlinear process responsible for transferring the
wave energy of beam-driven plasma oscillations to the nonresonant
part of turbulent spectrum is the modulational instability. The goal
of this paper is to investigate how the dispersion of linear plasma
modes and the growth rate of the modulational instability pumped by
the monochromatic Langmuir wave are modified in a strongly
non-maxwellian plasma. Our interest to this problem is motivated by
the need to interpret the experimental information about the
intensity and the frequency spectrum of electromagnetic radiation
generated in the turbulent plasma near the second harmonic of plasma
frequency. Recent calculations of the radiation power
\cite{tim1,tim2} for these experiments were based on the well known
analytical model of strong plasma turbulence \cite{gal}. According
to this model, the typical wavenumber of modulational instability
determines the size of the energy-containing region of turbulent
spectrum, and the typical growth rate allows to estimate the
saturation level of turbulence energy. Thus, the study of the
modulational instability in a non-maxwellian plasma will clarify
whether superthermal electrons are able to modify the main
parameters of emitting part of the turbulent spectrum.

The linear analysis of Langmuir waves in an isotropic plasma with a
power-law momentum distribution in the framework of relativistic
kinetic theory has been applied recently to the solar wind plasma
near the orbit of the Earth \cite{pod}. In our paper, we study the
dispersion properties of much hotter plasma that is typical to the
laboratory beam-plasma experiments, and we focus on the validity of
various widely used approximations. Investigations of modulational
instability in relativistic plasmas with the Maxwell-J\"{u}ttner
distribution \cite{liu1,liu2} or in the kappa-distributed plasmas
with nonrelativistic superthermal tails \cite{liu3,rio} are also of
interest in last years. In contrast to these papers, we consider the
modulational instability of a Langmuir wave in an unmagnetized
plasma with the relativistic power-law momentum distribution of
plasma electrons.

In Section II we obtain numerical solutions of the linear dispersion
equation for the potential plasma waves of isotropic non-maxwellian
plasma and compare them to the results predicted by fluid and
kinetic approximations. In Section III, for the modulational
instability, we deduce the dispersion equation that takes into
account relativistic and kinetic effects for high-frequency
oscillations, and compare numerical solutions for unstable spectra
with the case of maxwellian plasma. In the concluding Section IV, we
formulate our main results.

\section{Linear dispersion of potential plasma oscillations}

\begin{figure}[htb]
\bc\includegraphics[width=260bp]{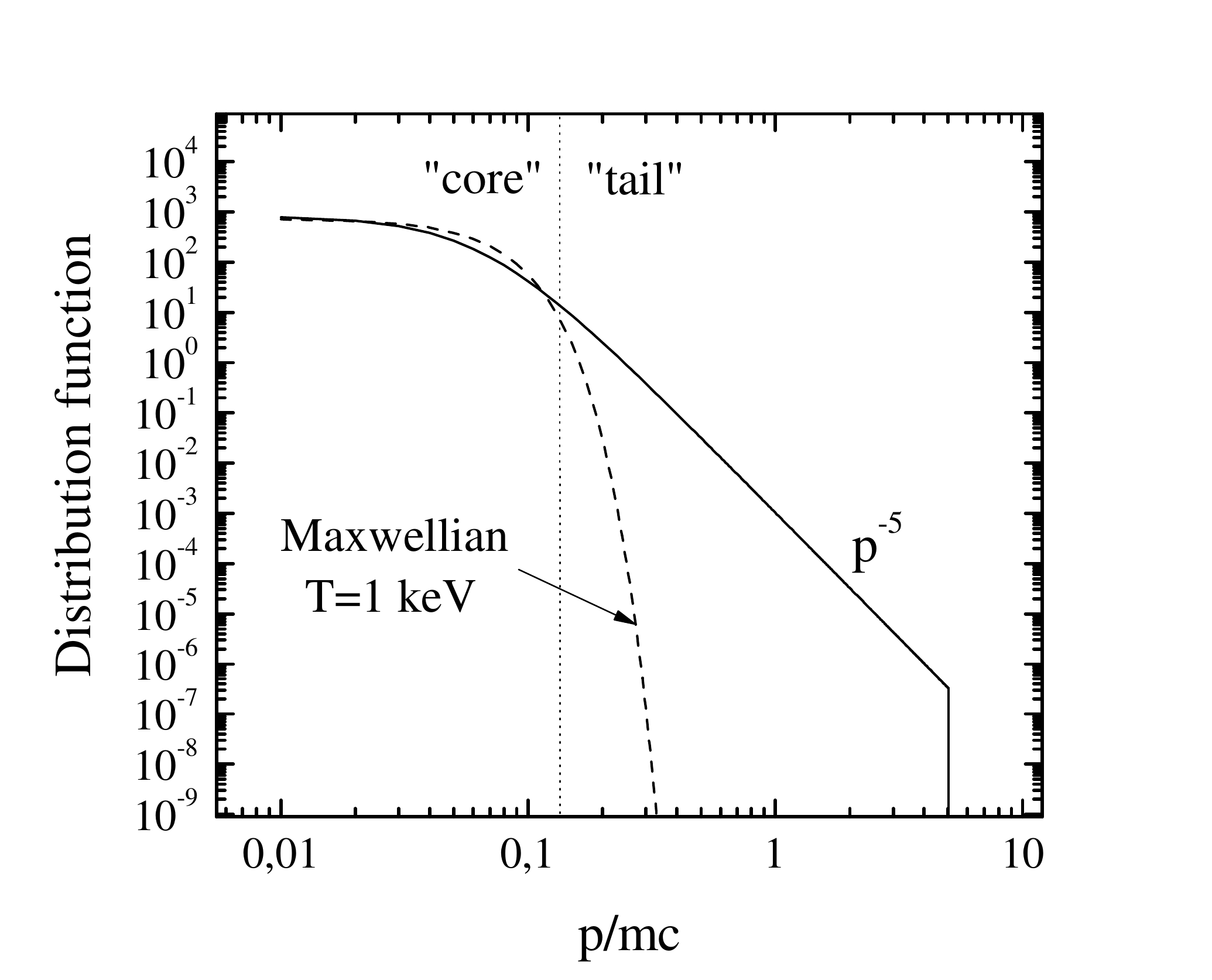} \ec \caption{Momentum
distribution of plasma electrons $f({\bf p})$ for parameters typical
to beam-plasma experiments at the GOL-3 facility.}\label{r1}
\end{figure}

Let us analyze the proper waves of non-maxwellian plasma assuming
cold ions and using the following distribution function for
electrons:
\begin{equation}\label{e1}
    f({\bf p})=\frac{C_0}{4 \pi} \frac{H(p_h-p)}{(p^2+\Delta p^2)^{5/2}},
\end{equation}
where $H(p_h-p)$ is the Heaviside step function, $p_h$ is the
threshold momentum, above which there are no electrons in the
plasma, $\Delta p$ is the typical momentum spread of plasma
electrons, and $C_0=3\Delta p^2/\sin^3 (p_h/\Delta p)$ is the
coefficient corresponding to the normalization $\int f d{\bf p}=1$.
For the parameters $\Delta p=0.066 m_e c$ and $p_h=5 m_e c$ ($m_e$
is the rest mass of electons, $c$ is the speed of light), this
function is best suited for the description of real electron
distribution observed in beam-plasma experiments at the GOL-3
facility. The feature of this distribution function is that, for
small momenta $p<\tilde{p}=2\Delta p$, it does not differ greatly
from the maxwellian function with the temperature $T=1$ keV  and,
for large momenta $p>\tilde{p}$, it has a power-law tail decreasing
so slowly that most of the plasma kinetic energy is concentrated in
superthermal electrons (fig. \ref{r1}).
The tail component of this particular distribution ($p>\tilde{p}$)
contains 28\% of the particles and 89\% of the kinetic energy; the
effective temperature of the whole distribution is
\begin{equation}\label{e2}
    T_{eff}=\frac{2}{3} \int m_e c^2 (\gamma-1) f d{\bf p}=6.5\,
    \mbox{keV},
\end{equation}
whereas the temperature of core electrons with the relative density
$n_c=72\%$ turns out to be much smaller:
\begin{equation}\label{e3}
    T_c=\frac{2}{3 n_c}\int\limits_{p<\tilde{p}} m_e c^2 (\gamma-1) f d{\bf p}=0.97\,
    \mbox{keV}
\end{equation}
($\gamma$ is the relativistic factor).

It is well known that dispersion laws of Langmuir and ion-acoustic
waves in the maxwellian plasma have the following dimensionless
forms:
\begin{equation}\label{e4}
    \omega_{\ell}\simeq 1+\frac{3}{2}k^2 T,
\end{equation}
\begin{equation}\label{e5}
    \omega_s\simeq \sqrt{\frac{m_e}{m_i}}\frac{k T^{1/2}}{\sqrt{1+k^2 T}}.
\end{equation}
Here frequencies are expressed in units of the plasma frequency
$\omega_p=(4\pi e^2 n/m_e)^{1/2}$, where $e$ is the electron charge
and $n$ is the plasma density, wavenumbers are measured in units of
$\omega_p/c$, and the electron temperature in units of $m_e c^2$.
These formulas have a simple hydrodynamic interpretation, according
to which the electron pressure gradient $\nabla \delta p=\gamma_e
T\nabla \delta n$ drives low-frequency acoustic oscillations of ions
($\gamma_e=1$) and creates an additional force for high-frequency
Langmuir oscillations of electrons ($\gamma_e=3$). To adapt these
insights to the case of non-maxwellian plasma, we should clarify
what temperature ($T_{eff}$ or $T_c$) is more appropriate for the
correct description of wave dispersion.

As we shall show later, the same expressions for real parts of wave
frequencies can be derived from the exact kinetic theory, if we use
the hydrodynamic approximation for Langmuir waves and kinetic
approximation for ion-acoustic waves. From the nonrelativistic
theory one can conclude that, for Langmuir oscillations, the role of
temperature is played by
\begin{equation}\label{e6}
    T_{\ell}=\frac{1}{3}\int v^2 f d{\bf
    p}=\left\langle\frac{v^2}{3}\right\rangle,
\end{equation}
whereas, for ion-acoustic oscillations, the temperature is defined
in another way
\begin{equation}\label{e7}
    T_s=\frac{1}{<1/v^2>}.
\end{equation}
In the maxwellian plasma with the temperature $T$, these values are
equal, $T_{\ell}=T_s=T$, but even from these simple formulas one can
see that relative contributions of superthermal electrons to
$T_{\ell}$ and $T_s$ differ substantially. In a non-maxwellian
plasma, the energetic tail component should lead to the significant
increase of $T_{\ell}$, since it is proportional to the averaged
kinetic energy of particles, and should not contribute to the value
$T_s$, in which the inverse energy should be averaged. Thus, in the
nonrelativistic case, superthermal electrons increase the thermal
correction to the Langmuir wave frequency that is determined by the
effective temperature of the whole distribution $T_{eff}$ and have
almost no impact on ion-acoustic waves, for which the thermal spread
of core electrons $T_c$ is more suitable to the role of temperature.

Let us now find out how these results change, if the tail of
superthermal electrons extends to relativistic energies. First we
will derive dispersion laws $\omega_r(k)$ and damping rates
$\Gamma(k)$ of potential plasma waves in the hydrodynamic and
kinetic limits, and then investigate applicability of these
approximations by comparing approximate results with numerical
solutions of the exact dispersion relation.

For slowly damping waves of isotropic plasma, the dispersion
equation
\begin{equation}\label{e8}
    \varepsilon_{\parallel}=1+\frac{1}{k^2}\int \frac{{\bf k}\cdot \partial
    f/ \partial {\bf p}}{\omega - {\bf k} {\bf v}} d{\bf
    p}-\frac{m_e/m_i}{\omega^2}=0
\end{equation}
can be reduced to the form
$$\mbox{Re}\, \varepsilon_{\parallel}
(\omega_r-i \Gamma,k)+i \mbox{Im}\, \varepsilon_{\parallel}
(\omega_r,k)=0,$$ where real and imaginary parts of dielectric
permittivity are defined by the integrals
\begin{equation}\label{e9}
    \mbox{Re}\, \varepsilon_{\parallel}=1+\frac{1}{k^2} VP\int \frac{{\bf k} {\bf
    v}}{v}  \frac{\partial f/ \partial p}{\omega-{\bf k} {\bf
    v}} d {\bf p} -\frac{m_e/m_i}{\omega^2},
\end{equation}
\begin{equation}\label{e10}
    \mbox{Im}\, \varepsilon_{\parallel}=-\frac{\pi}{k^2}\int \frac{{\bf k} {\bf
    v}}{v} \frac{\partial f}{\partial p} \delta(\omega-{\bf k} {\bf
    v}) d {\bf p}.
\end{equation}
The dispersion law in this case is found from the equation
\begin{equation}\label{e11}
     \mbox{Re}\, \varepsilon_{\parallel} (\omega_r, k)=0,
\end{equation}
and the damping rate is calculated as
\begin{equation}\label{e12}
     \Gamma=\left(\frac{\mbox{Im}\, \varepsilon_{\parallel}}{\partial \mbox{Re}\, \varepsilon_{\parallel}
/\partial \omega}\right)_{\omega=\omega_r}.
\end{equation}
We can calculate the electron contribution to $\mbox{Re}\,
\varepsilon_{\parallel}$ using the hydrodynamic approximation
($\omega\gg{\bf k} {\bf v}$) for Langmuir waves and the kinetic
approximation ($\omega\ll{\bf k} {\bf v}$) for ion-acoustic waves.
It means that, in the former case, we can use the expansion
\begin{equation}\label{e13}
    \frac{1}{\omega-{\bf k} {\bf v}}=\frac{1}{\omega}\left(1+
    \frac{{\bf k} {\bf v}}{\omega}+\frac{({\bf k} {\bf v})^2}{\omega^2}+
    \frac{({\bf k} {\bf v})^3}{\omega^3}...\right),
\end{equation}
and in the latter case
\begin{equation}\label{e14}
\frac{1}{\omega-{\bf k} {\bf v}}=-\frac{1}{{\bf k} {\bf v}}+...
\end{equation}
$\mbox{Im}\, \varepsilon_{\parallel}$ is determined for both types
of waves by the unified expression:
\begin{equation}\label{e15}
\mbox{Im}\, \varepsilon_{\parallel}=2 \pi^2
\frac{\omega}{k^3}\left[\gamma_0^2
f(p_0)+2\int\limits_{p_0}^{\infty} p f dp\right],
\end{equation}
where $p_0$ is the momentum of electrons getting into Cherenkov
resonance with the wave,
$$p_0=\frac{\omega_r/k}{\sqrt{1-\omega_r^2/k^2}},$$
and $\gamma_0=\sqrt{1+p_0^2}$ is the corresponding relativistic
factor. Thus, the real frequency of Langmuir wave is governed by the
equation
\begin{equation}\label{e16}
    1-\frac{\omega_0^2}{\omega_r^2}-3\frac{k^2
    T_{\ell}}{\omega_r^4}=0,
\end{equation}
where
\begin{equation}\label{e17}
    \omega_0^2=\int \frac{f}{\gamma}\left(1-\frac{v^2}{3}\right) d{\bf
    p}=\left\langle\frac{1-v^2/3}{\gamma}\right\rangle,
\end{equation}
\begin{equation}\label{e18}
    T_{\ell}=\left\langle\frac{v^2}{3\gamma} \left(1-\frac{3}{5}v^2\right)\right\rangle,
\end{equation}
and the real frequency of ion-acoustic wave is determined by
\begin{equation}\label{e19}
    1+\frac{1}{k^2 T_s}-\frac{m_e/m_i}{\omega_r^2}=0,
\end{equation}
where
\begin{equation}\label{e20}
    T_s^{-1}=\left\langle\frac{1+v^2}{pv}\right\rangle.
\end{equation}
In the nonrelativistic limit, $\omega_0^2\rightarrow 1$ and
calculation of electron temperature really reduces to averaging
either the kinetic energy $T_{\ell}\rightarrow <v^2/3>$ or the
reciprocal value $T_s^{-1}\rightarrow <1/v^2>$. From these formulas,
it is seen that relativistic effects manifest themselves in
decreasing the plasma frequency ($\omega_0^2<1$) due to weighting of
tail electrons. Moreover, the temperature $T_{\ell}$ in the
relativistic case ceases to be proportional to the total kinetic
energy and can no longer be identified with the value $T_{eff}=<2
(\gamma -1)/3>$. Indeed, the expression averaged in $T_{\ell}$
(\ref{e18}) differs from the value $2 (\gamma -1)/3$ by the factor
\begin{equation}\label{e21}
    F=\frac{(\gamma +1)}{5 \gamma^3} \left(1+\frac{3}{2
    \gamma^2}\right),
\end{equation}
which substantially reduces the relative contribution of fast
electrons to the thermal correction to the Langmuir wave frequency.
As to the temperature $T_s$, its calculation for both low $\gamma\ll
1$ and high $\gamma\gg 1$ energies reduces to averaging the value
that is inversely proportional to the energy. It means that
relativistic effects do not change the earlier conclusion that
superthermal electrons do not contribute to the electron pressure
perturbation driving low-frequency acoustic oscillations. For the
distribution function (\ref{e1}), the temperature $T_{\ell}$
accounting for the relativistic effects appears to be twice smaller
than the effective temperature of the whole distribution
($T_{\ell}=0.56\, T_{eff}$) and the temperature $T_s$  remains close
to the temperature of low-energy core electrons ($T_s=1.15\, T_c$)
as in the nonrelativistic case.

\begin{figure}[htb]
\bc\includegraphics[width=210bp]{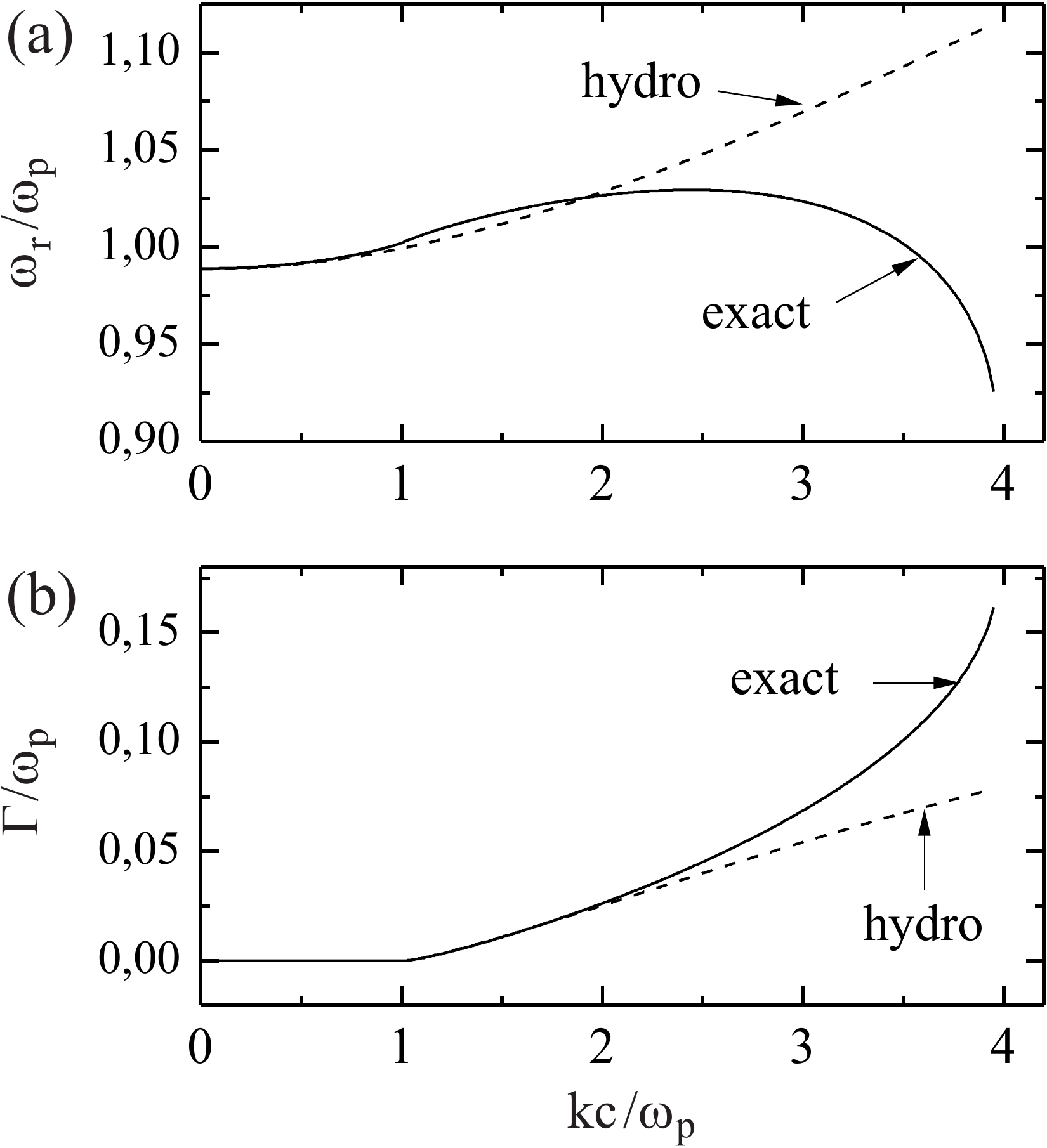} \ec \caption{The
dispersion (a) and Landau damping (b) of Langmuir waves in the
non-maxwellian plasma.}\label{r2}
\end{figure}

\begin{figure}[htb]
\bc\includegraphics[width=210bp]{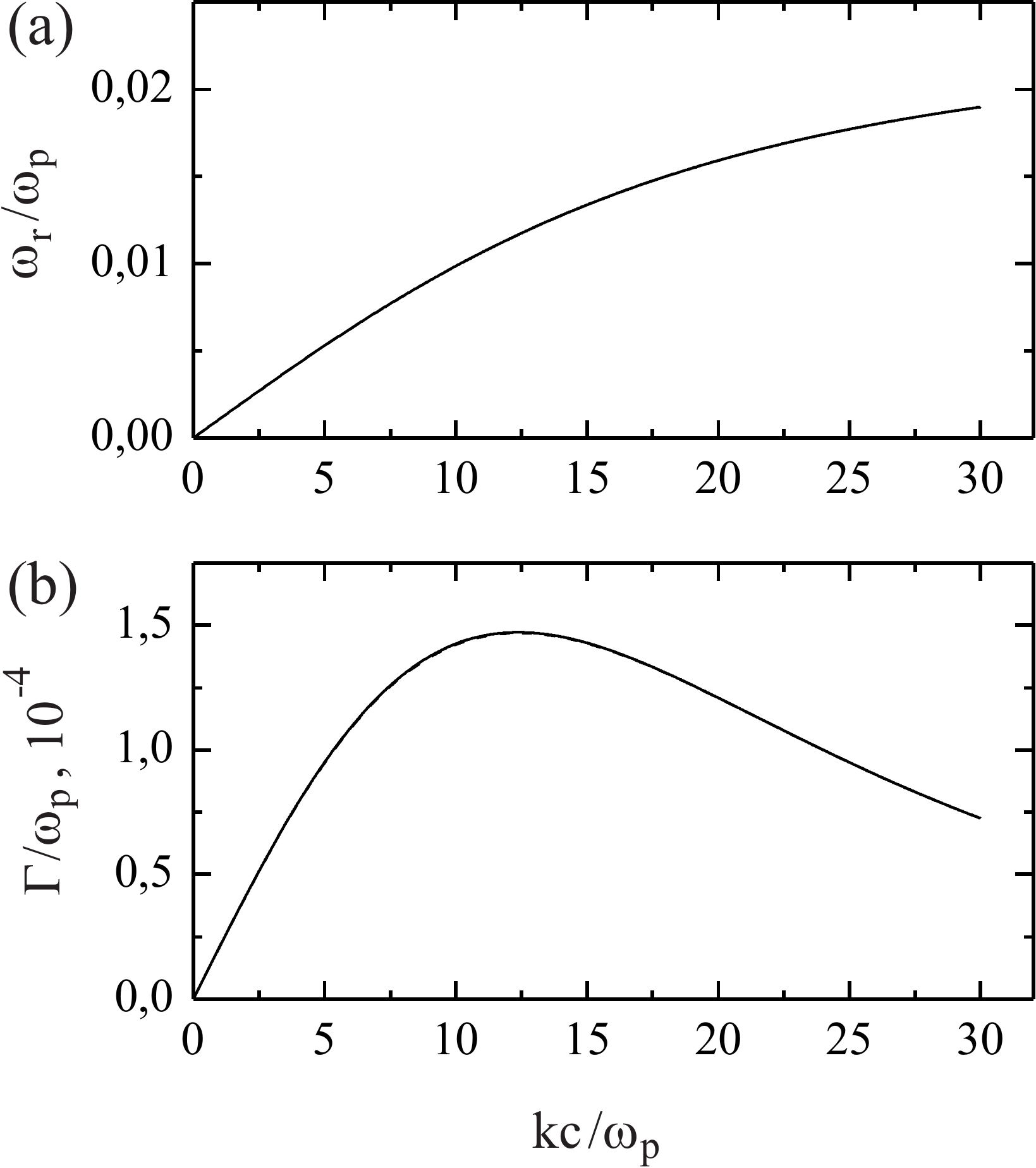} \ec \caption{The
dispersion (a) and Landau damping (b) of ion-acoustic waves in the
non-maxwellian plasma (numerical solutions of equation (\ref{e22})
coincide with the results of kinetic approximation).}\label{r3}
\end{figure}

The exact dispersion relation for the damping plasma modes can be
written in the form
\begin{multline}\label{e22}
    \varepsilon_{\parallel}(\omega, k)=1-\frac{m_e/m_i}{\omega^2}+\frac{4
    \pi}{k^2}\int\limits_0^{\infty} \frac{f}{\gamma} (\gamma^2+p^2)
    dp \\
    -2\pi \frac{\omega}{k^3}\int\limits_0^{\infty}\gamma^2 \frac{\partial f}{\partial p}
    \ln \left(\frac{\omega-k v}{\omega+k v}\right)d p+i \mbox{Im}
    \varepsilon_{\parallel}=0,
\end{multline}
where the imaginary part of dielectric permittivity is still defined
by (\ref{e15}) and the function $\ln(...)$ means the principal
branch of complex logarithm. For Langmuir waves, the comparison of
numerical solution of this equation $\omega=\omega_r-i\Gamma$ with
the results of the hydrodynamic approximation
\begin{equation}
    \omega_r=\left(\frac{\omega_0^2+\sqrt{\omega_0^4+12 k^2
    T_{\ell}}}{2}\right)^{1/2},
\end{equation}
\begin{multline}
    \Gamma=\frac{\pi \omega_r^4 C_0 }{6 k^3 \left(\omega_0^2+6k^2
    T_{\ell}/\omega_r^2\right) (p_0^2+\Delta p^2)^{5/2}} \times \\
    \left[\frac{3\gamma_0^2}{2}+p_0^2+\Delta p^2
    -\frac{(p_0^2+\Delta p^2)^{5/2}}{(p_h^2+\Delta
    p^2)^{3/2}}\right],
\end{multline}
is shown in Fig. \ref{r2}.
It is seen that the approximate solution for the parameters of
interest is adequate in the long-wavelength ($k<2$) part of the
spectrum only. Oscillations with shorter wavelength falls in the
region of strong Landau damping and demonstrate the anomalous
dispersion ($\partial \omega_r/\partial k<0$). For ion-acoustic
waves, on the contrary, the kinetic approximation (\ref{e19}) turns
out to be so accurate that both real and imaginary parts of complex
frequency fit the numerical solution of exact dispersion equation in
the wide range of wavenumbers (fig. \ref{r3}).

\section{Modulational instability}

Let us consider stability of the monochromatic Langmuir wave
\begin{equation}
    {\bf E}(t,{\bf r})=\frac{1}{2}\left({\bf E_0} e^{\ds -i \omega_k t +i {\bf k}{\bf r}}+c.c.\right)
\end{equation}
with respect to the coupled system of low-frequency and
high-frequency modulational perturbations
\begin{multline}
    {\bf \delta E}(t,{\bf r})=\frac{{\bf\delta E}}{2} e^{\ds -i \Omega t +i {\bf q}{\bf
    r}}+ \frac{{\bf\delta E_{+}}}{2} e^{\ds -i \omega_{+} t +i {\bf k_{+}}{\bf
    r}}\\
+\frac{{\bf\delta E_{-}}}{2} e^{\ds -i \omega_{-} t +i {\bf
k_{-}}{\bf r}}+c.c.,
\end{multline}
where
$$ \omega_{\pm}=\omega_k\pm \Omega, \qquad {\bf k_{\pm}}={\bf k}\pm {\bf q}.$$
In the plasma with cold ions and arbitrarily distributed electrons,
equations for the amplitudes of potential high-frequency satellites
can be written in the following form:
\begin{multline}
    \varepsilon_{+}\delta E_{+}=\frac{e^2 E_0}{4 k k_{+}} \times \\
    \left[\frac{E_0^{\ast} \delta E_{+}}{k k_{+}}\left(\frac{G_1^+ G_4}{q^2 \varepsilon_q}-G_2^+\right)
    +\frac{E_0 \delta E_{-}^{\ast}}{k k_{-}}\left(\frac{G_1^+ G_5}{q^2
    \varepsilon_q}-G_3^+\right)\right],
\end{multline}
\begin{multline}
    \varepsilon_{-}^{\ast}\delta E_{-}^{\ast}=\frac{e^2 E_0^{\ast}}{4 k k_{-}} \times \\
    \left[\frac{E_0^{\ast} \delta E_{+}}{k k_{+}}\left(\frac{G_1^- G_4}{q^2 \varepsilon_q}-G_2^-\right)
    +\frac{E_0 \delta E_{-}^{\ast}}{k k_{-}}\left(\frac{G_1^- G_5}{q^2
    \varepsilon_q}-G_3^-\right)\right],
\end{multline}
where
\begin{align}
G_1^{\pm}&=4 \pi e^2 \int \frac{d^3 p}{\omega_{\pm}-{\bf
k_{\pm}}{\bf v}}\times \nonumber \\ &\left[{\bf k}\cdot
\partial_{\bf p}\left(\frac{{\bf q}\cdot
\partial_{\bf p} f}{\Omega-{\bf q}{\bf v}}\right)\pm {\bf q}\cdot
\partial_{\bf p}\left(\frac{{\bf k}\cdot
\partial_{\bf p} f}{\omega_k-{\bf k}{\bf v}}\right)\right], \\
G_2^{\pm}&=4 \pi e^2 \int d^3 p\frac{{\bf k}\cdot
\partial_{\bf p}}{\omega_{\pm}-{\bf k_{\pm}}{\bf
v}}\times \nonumber \\ &\left[\frac{{\bf k_+}\cdot
\partial_{\bf p}}{\Omega-{\bf q}{\bf v}}\left(\frac{{\bf k}\cdot
\partial_{\bf p} f}{\omega_k-{\bf k}{\bf v}}\right)- \frac{{\bf k}\cdot
\partial_{\bf p}}{\Omega-{\bf q v}}\left(\frac{{\bf k_+}\cdot
\partial_{\bf p} f}{\omega_+-{\bf k_+}{\bf v}}\right)\right], \\
G_3^{\pm}&=4 \pi e^2 \int d^3 p\frac{{\bf k}\cdot
\partial_{\bf p}}{\omega_{\pm}-{\bf k_{\pm}}{\bf
v}}\times \nonumber \\ &\left[\frac{{\bf k}\cdot
\partial_{\bf p}}{\Omega-{\bf q}{\bf v}}\left(\frac{{\bf k_-}\cdot
\partial_{\bf p} f}{\omega_- -{\bf k_-}{\bf v}}\right)- \frac{{\bf k_-}\cdot
\partial_{\bf p}}{\Omega-{\bf q v}}\left(\frac{{\bf k}\cdot
\partial_{\bf p} f}{\omega_k-{\bf k}{\bf v}}\right)\right], \\
G_4&=4 \pi e^2 \int \frac{d^3 p}{\Omega-{\bf q}{\bf v}}\times
\nonumber \\ &\left[{\bf k_+}\cdot
\partial_{\bf p}\left(\frac{{\bf k}\cdot
\partial_{\bf p} f}{\omega_k-{\bf k}{\bf v}}\right)- {\bf k}\cdot
\partial_{\bf p}\left(\frac{{\bf k_+}\cdot
\partial_{\bf p} f}{\omega_+-{\bf k_+}{\bf v}}\right)\right], \\
G_5&=4 \pi e^2 \int \frac{d^3 p}{\Omega-{\bf q}{\bf v}}\times
\nonumber \\ &\left[{\bf k}\cdot
\partial_{\bf p}\left(\frac{{\bf k_-}\cdot
\partial_{\bf p} f}{\omega_--{\bf k_-}{\bf v}}\right)- {\bf k_-}\cdot
\partial_{\bf p}\left(\frac{{\bf k}\cdot
\partial_{\bf p} f}{\omega_k-{\bf k}{\bf v}}\right)\right],
\end{align}
$$ \varepsilon_{\pm}=\varepsilon_{\parallel}(\omega_{\pm}, {\bf k_{\pm}}),
\quad \varepsilon_q=\varepsilon_{\parallel}(\Omega, {\bf q}).$$ In
the limiting cases $\Omega\ll {\bf q v}$ and $\omega_{k,\pm}\gg
{\bf k_{\pm} v}$, all integrals  $G_i$ reduce to averaging
$<1/v^2>$. Thus, the main contribution to these integrals comes
from low-energy core electrons, the typical velocity of which is
really much smaller than  phase velocities of high-frequency
satellites and the speed of light. It allows us to neglect
relativistic effects in calculations of $G_i$ and transform the
equations to the well-known form \cite{khak}
\begin{align}
\varepsilon_+ \delta E_+ &= -\frac{E_0 \cos \theta_+}{16 \pi n T_s}
\frac{\varepsilon_q^i}{\varepsilon_q} \left(E_0^{\ast} \delta E_+
\cos \theta_+ + E_0 \delta E_-^{\ast} \cos \theta_-\right), \\
\varepsilon_-^{\ast} \delta E_-^{\ast} &= -\frac{E_0^{\ast} \cos
\theta_-}{16 \pi n T_s} \frac{\varepsilon_q^i}{\varepsilon_q}
\left(E_0^{\ast} \delta E_+ \cos \theta_+ + E_0 \delta E_-^{\ast}
\cos \theta_-\right),\\
&\varepsilon_q^i=-\omega_{pi}^2/\Omega^2, \qquad \cos
\theta_{\pm}=\frac{{\bf k k_{\pm}}}{k k_{\pm}},
\end{align}
The specific features of non-maxwellian plasma in these equations
manifest themselves only in changing of temperature notation from
$T$ to $T_s$ and in accurate calculation of permittivities
$\varepsilon_{\pm}$ in the framework of relativistic kinetic
theory.

Thus, the modulational instability in a non-maxwellian plasma can
be described by the following equation:
\begin{equation}\label{e36}
    \Omega^2-\Omega_s^2=W\Omega_s^2 \left[\frac{\cos^2 \theta_+}{\varepsilon_+}
    +\frac{\cos^2 \theta_-}{\varepsilon_-^{\ast}}\right]
\end{equation}
where
$$\Omega_s^2=\frac{m_e}{m_i}\frac{q^2 T_s}{1+q^2 T_s} $$
corresponds to the linear dispersion of ion-acoustic waves, and the
parameter
$$ W=\frac{|E_0|^2}{16 \pi n T_s}$$
determines the relative energy of the Langmuir pump wave. Since we
look for unstable solutions of (\ref{e36}) $\Omega=\mbox{Re}\,\Omega
+i \mbox{Im}\,\Omega$ with $\mbox{Im}\,\Omega>\mbox{Im}\,\omega_k$,
the last term in the definition of $\varepsilon_{\parallel}$,
appearing as a result of analytical continuation of this function to
the lower half-plane of $\omega$, must be omitted.

\begin{figure}[htb]
\bc\includegraphics[width=220bp]{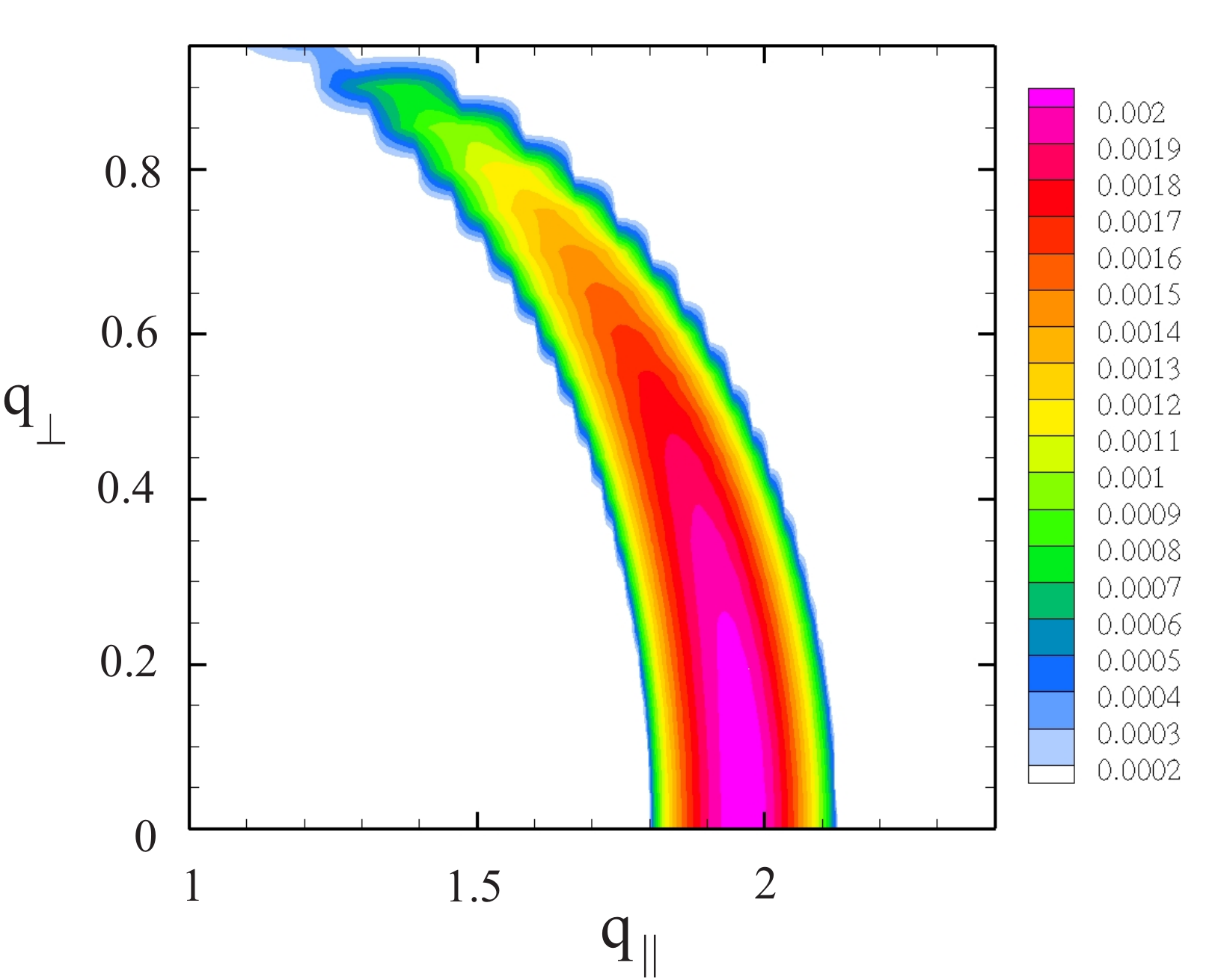} \ec \caption{The
growth-rate map for the modulational instability pumped by a
monochromatic Langmuir wave.}\label{r4}
\end{figure}

\begin{figure}[htb]
\bc\includegraphics[width=210bp]{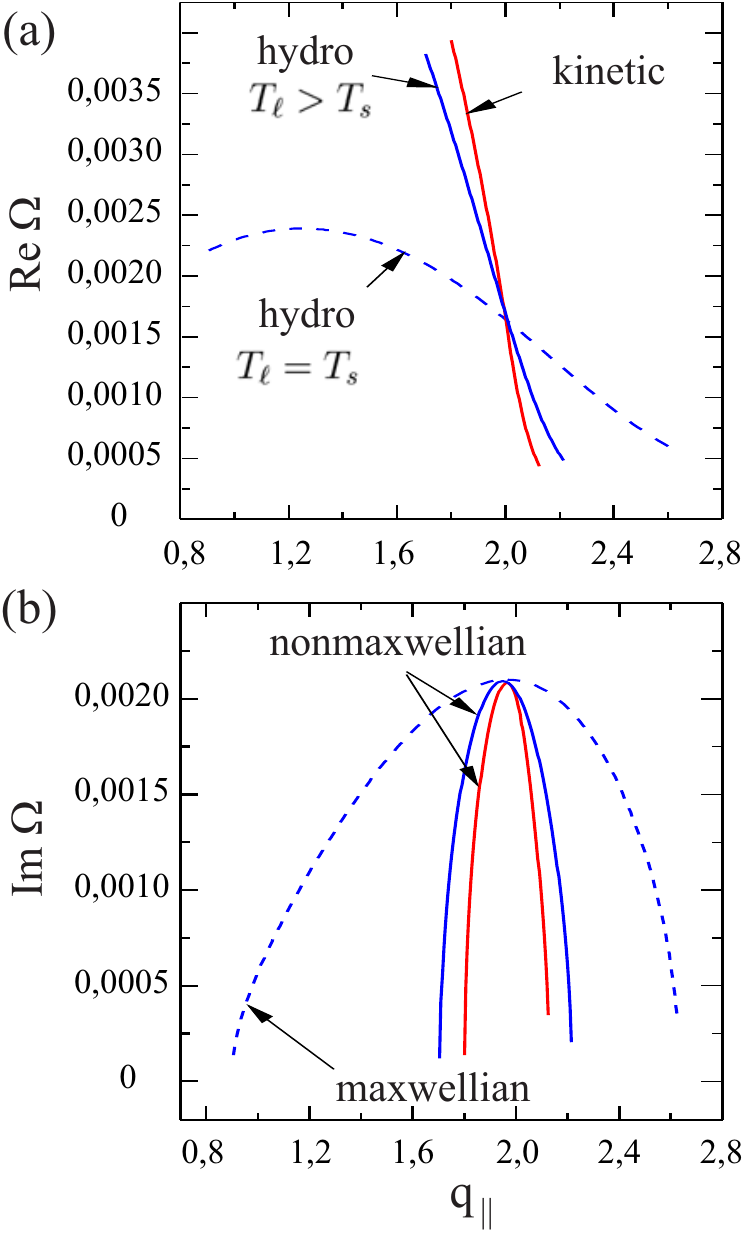} \ec \caption{The real (a)
and imaginary (b) parts of the frequency of longitudinally
propagating modulational perturbations  (the fluid approximation for
the maxwellian plasma with $T_{\ell}=T_s$ -- dashed line, the fluid
approximation for the non-maxwellian plasma with the increased
temperature $T_{\ell}$ -- blue solid line, numerical solution of
(\ref{e36}) with the exact kinetic expressions for
$\varepsilon_{\pm}$ -- red solid line).}\label{r5}
\end{figure}

For example, let us find a numerical solution of the dispersion
relation (\ref{e36}) in the case when the role of long-wavelength
pump is played by the undamped Langmuir wave with the wavenumber
$k=1$, the frequency $\omega_k=1.00203$ and the energy $W=0.01$.
Fig. \ref{r4} shows that
the growth rate of modulational instability
$\mbox{Im}\,\Omega(q_{\perp}, q_{\parallel})$ reaches the maximum
for the perturbations propagating along ${\bf k}$. Since this
maximum falls in the region of rather small wavenumbers
$q_{\parallel}\simeq 2$, it is reasonable to estimate how accurately
the unstable spectrum can be described by the fluid approach, in
which thermal corrections to the dispersion of high-frequency
oscillations depends on the increased effective temperature
$T_{\ell}>T_s$. In this limit, the functions $\varepsilon_{\pm}$ in
the dispersion equation should have the simple form
\begin{equation}
    \varepsilon_{\pm}=1-\frac{1+3 k_{\pm}^2
    T_{\ell}}{\omega_{\pm}^2}.
\end{equation}
The unstable spectrum of longitudinally propagating modulational
perturbations ($q_{\perp}=0$) in comparison with the case of
maxwellian plasma ($T_{\ell}=T_s$) is shown in Fig. \ref{r5}.
It is seen that the increase in the temperature $T_{\ell}$ with
respect to $T_s$ in the hydrodynamic limit results in almost the
same narrowing of the spectrum of modulational instability (by the
factor $T_{\ell}/T_s$) that is observed in the case of exact kinetic
description of Langmuir waves. Another important result is that the
appearance of the tail component in the electron distribution does
not change neither the maximum of the growth rate nor its position
in the wavenumber space. It means that we can estimate the typical
gtowth rate and the corresponding wavenumber of modulational
instability in the non-maxwellian plasma without the detailed study
of the real distribution function. For this purpose, we can restrict
ourselves by the case of maxwellian plasma with the temperature
corresponding to the temperature of core electrons.

\section{Summary}

In the present study we have calculated dispersion laws and damping
rates of Langmuir and ion-acoustic oscillations in the case, when an
isotropic plasma has the energetic tail of superthermal electrons
typical to the beam-plasma experiments at the multi-mirror trap
GOL-3. The specific feature of such distribution is that most of the
kinetic energy is concentrated in a small population of fast
electrons. The concept of temperature in this case loses its usual
meaning, since the energy spread of the bulk of electrons appears to
be much smaller than the total kinetic energy. Nevertheless, the
dispersion of linear Langmuir waves in the long-wavelength region
can be correctly described by the fluid approach, in which the
effective temperature $T_{\ell}$ exceeds substantially the
temperature of core electrons $T_c$, but, due to relativistic
effects, does not reach the temperature of the whole distribution.
Ion-acoustic waves, in turn, are shown to be well described by the
kinetic approximation with the temperature of low-energy core
electrons $T_s\approx T_c$.

Modifications in the linear dispersion of high-frequency waves in a
non-maxwellian plasma should have an impact on the build-up
efficiency of modulational instability playing the important role in
the scenarios of strong plasma turbulence. We have solved
numerically the dispersion equation for modulational perturbations,
in which both kinetic and relativistic effects are taken into
account. It is shown that the unstable spectrum in the
non-maxwellian plasma narrows significantly without changing in the
growth rate and wavelength of the most unstable perturbation. It is
also found that the same trend is qualitatively reproduced in the
case when high-frequency waves are described by the fluid
approximation with the increased effective temperature
$T_{\ell}>T_s$.

Independence of the typical growth rate and the wavelength of
modulational instability on distribution details of superthermal
electrons allows to prove the possibility of applying the model of
strong Langmuir turbulence \cite{tim1}, used for calculation of
electromagnetic plasma emission, to the case of realistic
non-maxwellian plasmas typical to laboratory beam-plasma
experiments.

The author thanks K.V.Lotov for useful discussions. The study was
supported by The Ministry of education and science of Russia
(projects 14.B37.21.0750, 14.B37.21.1178 and 14.B37.21.0784),
Russian Foundation of Basic Research (grants 12-02-31696,
11-02-00563), grant of RF Government 11.G34.31.0033 and President
grants SP-1289.2012.1 and NSh-5118.2012.2.

\end{document}